\pdfoutput=1
\documentclass[journal=jpcl,manuscript=article,layout=twocolumn]{achemso}

\usepackage{graphicx}
\usepackage{dcolumn}
\usepackage{bm}
\usepackage{dsfont}
\usepackage{amsmath}
\usepackage{physics}
\usepackage{color}
\usepackage{soul}
\usepackage{url}
\usepackage[hidelinks]{hyperref}
\usepackage{scalerel}

\title{ On-the-Fly Cavity–Molecular Dynamics of Vibrational Polaritons }

\author{Sachith Wickramasinghe}
\affiliation{Department of Chemistry, Texas A\&M University, College Station, Texas 77843, USA}
\altaffiliation{equal contribution}

\author{Amirhosein Amini}
\affiliation{Department of Chemistry, Texas A\&M University, College Station, Texas 77843, USA}
\altaffiliation{equal contribution}

\author{Arkajit Mandal}
\email{mandal@tamu.edu}
\affiliation{Department of Chemistry, Texas A\&M University, College Station, Texas 77843, USA}

\begin{document} 

\begin{abstract}
{\footnotesize
  In this work, we combine the density functional tight-binding (DFTB) approach with a light-matter Hamiltonian beyond the long-wavelength approximation to propagate the dynamics of vibrational polaritons formed by coupling molecular vibrations to confined radiation inside a Fabry-P\'{e}rot optical cavity. Here, we develop a parallelized propagation scheme with lightweight inter-CPU communication by exploiting the sparse nature of the light-matter interactions  in the real space representation. We find that the computationally expensive Born charges required for our propagation can be replaced with the computationally inexpensive Mulliken charges to obtain qualitatively accurate linear spectra especially when the nonlinearity (arising from molecular vibrations) of the light-matter interaction term  is not substantial. However, the same approach may not be suitable to be used for studying cavity modification of energy transport or chemical dynamics as this approximation leads to spurious heating of the light-matter hybrid system. We demonstrate the utility of this on-the-fly approach to compute angle resolved polaritonic spectra of water. We implement our approach as an open-source computational package, {\it CavOTF}, which is available on GitHub.}
  \begin{figure}
    \centering
    \includegraphics[width=1.0\linewidth]{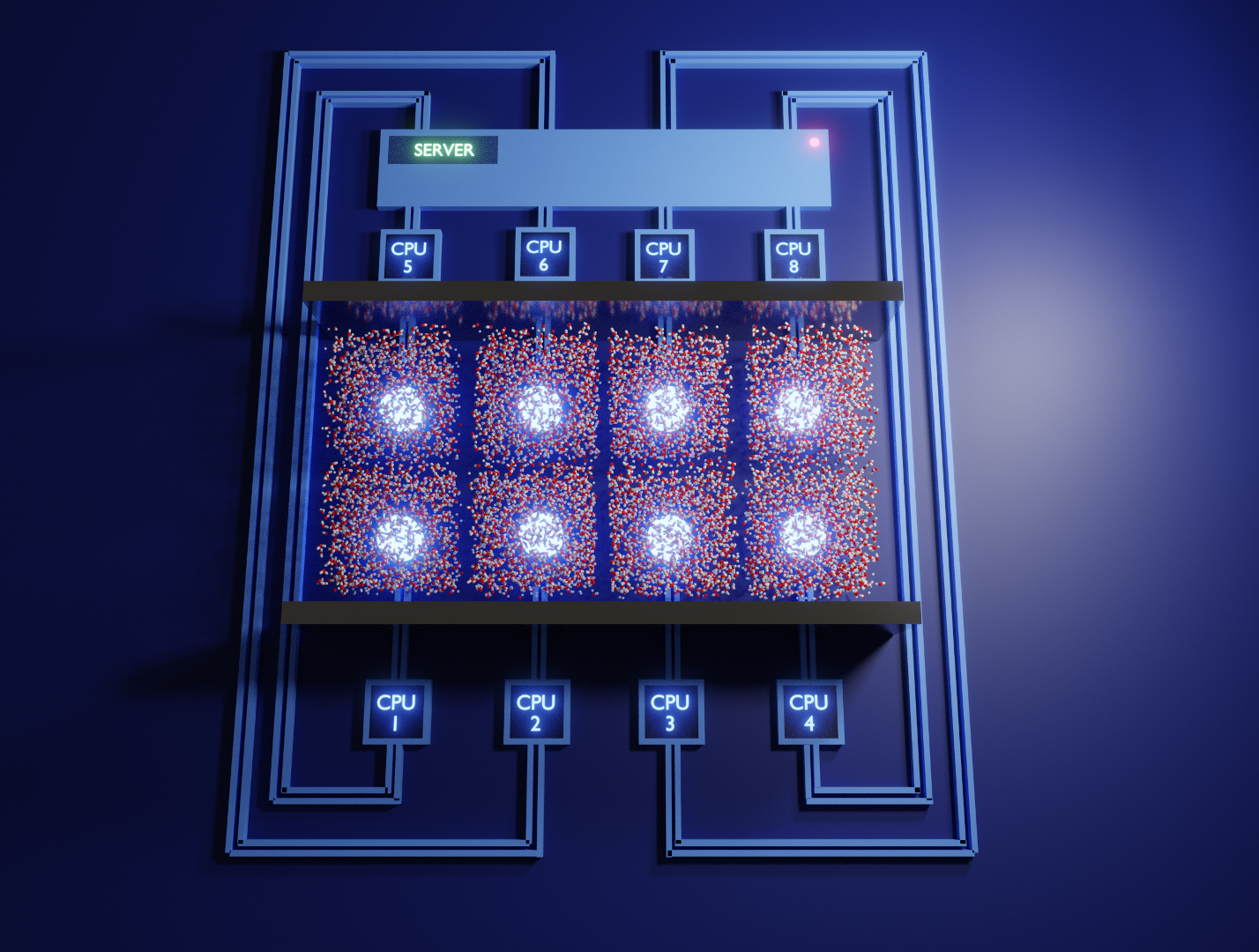} 
   \end{figure}

\end{abstract}
\maketitle
{\footnotesize
\section{Introduction}


A series of experiments have suggested that ground-state chemical reactions can be modified by coupling molecular vibrations to quantized vacuum radiation inside an optical cavity.~\cite{thomas2019Science, ahn2023Science, nagarajan2021JACS, guoxinyin2025Science, mandal2023ChemicalReviews, li2022Annualreview, ruggenthaler2023ChemicalReviews,campos2023jcp,lather2020improving}  In this light-matter coupling regime, namely the vibrational strong coupling (VSC) regime, molecular vibrational modes hybridize with the confined photonic modes to form vibro-polaritons which have been suggested to modify chemical dynamics in the absence of external illumination~\cite{kenacohen2019ACSS}.
Such couplings can lead to either enhancement~\cite{lather2019Wiley, nagarajan2021JACS, thomas2016Wiley} or suppression~\cite{ahn2023Science, nagarajan2021JACS, hirai2020Wiley} of bond-breaking processes (or produce no measurable change~\cite{nelson2024more, chen2024exploring, fidler2023ultrafast}). These effects are resonant, emerging when the cavity mode is tuned to a specific molecular vibration, thereby opening up the possibility to achieve chemical selectivity by tuning cavity frequency~\cite{thomas2016Wiley, vergauwe2019Wiley, lather2019Wiley,hirai2020Wiley,lather2020improving, lather2022Chemicalscience, Francisco2021}. Thus, VSC represents a fundamentally new and non-invasive approach to controlling chemistry, requiring no external pumping or chemical reagents; instead it harnesses the ever-present vacuum fluctuations of confined radiation. 




However, VSC remains far from a fully reliable or broadly applicable strategy for controlling ground-state chemical reactivity. This is because these remarkable experimental observations currently lack a clear connection to established chemical principles or molecular morphology, or even a semi-empirical framework to guide expectations does not yet exist. Significant challenges also persist in interpretation, particularly in disentangling genuine cavity-induced changes from other experimental factors as well as in the reproducibility of experimental results.~\cite{michon2024impact,wiesehan2021jcp, imperatore2021jcp} These limitations underscore the need for a clear microscopic understanding of cavity-modified ground-state chemical reactivity and the need for accurate modeling of light-matter interactions to unambiguously distinguish the role of cavity radiation in modifying ground-state chemical reactivity.\cite{lindoy2024Nanophotonics,imperatore2021jcp,wiesehan2021jcp}






Recent theoretical works that study a simple single-molecule-single-cavity-mode model system find resonant modification of chemical kinetics  under vibrational strong coupling.~\cite{lindoy2023NatureCommunications,vega2025jacs,ying2024Nanophotonics,ying2024CommunicationsMaterials, sun2023jpcletters} In contrast, theoretical works that consider an ensemble of molecules coupling to one cavity mode predict little or no change in chemical reactivity under similar conditions~\cite{campos2023jcp, lindoy2024Nanophotonics}. This discrepancy raises a fundamental question about what essential physics may be missing in these current theoretical descriptions. Likely limitations include the use of the long-wavelength and single-mode cavity approximations, as well as simplified molecular models that neglect vibrational anharmonicity and bond dissociation. Recently, Li and co-workers have introduced mesoscale simulation approaches that enable atomistic modeling of molecular ensembles within optical cavities.~\cite{li2021collective, ji2025selective, li2024mesoscale, li2022Naturecommunications} However, these methods typically rely on classical force fields, which are inherently nondissociative and thus unable to fully capture the chemical complexity, and reactive character intrinsic to real molecular systems.~\cite{sun2023jpcletters, wang2022jcp, wang2021ACSphotonics,schafer2022NatureCommunications, mondal2022jcp, liu202jpcC, li2021Naturecommunications, mandal2022jcp, campos2019Naturecommunications, xiang2020Science, brawley2025NatureChemistry, poh2023jpcc, yang2021jpcletters, Simpkiins2021, D4SC07053D, doi:10.1021/acs.jctc.4c00129}

\begin{figure*}[t]
    \centering
    \includegraphics[width=0.9\linewidth]{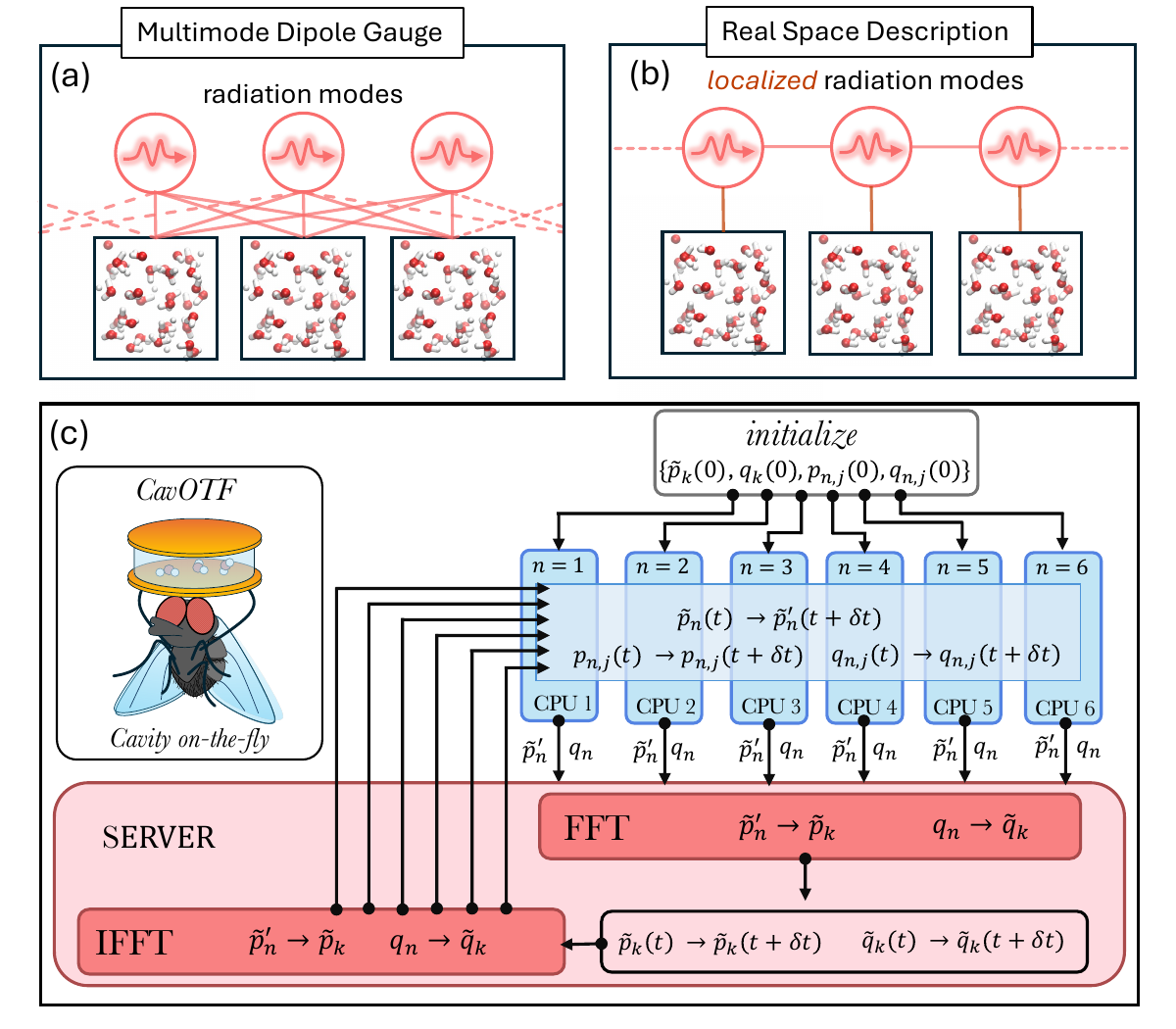} 
    \caption{\footnotesize Schematic illustration of light-matter interactions and the real-reciprocal used here for propagating vibro-polariton dynamics. Schematic illustration of light-matter interaction (a) in the standard multi-mode dipole gauge description and (b) in real space. (c) Schematic illustration of the algorithm used here and implemented in our computational package {\it CavOTF}.\cite{cavOTF}} 
    \label{Fig1}
\end{figure*}
 

In this work, we develop an on-the-fly molecular dynamics approach in which part of the Hamiltonian is evolved in real space, while the rest is propagated in reciprocal space. This strategy enables the development of a highly parallelized computational algorithm that requires only light inter-CPU communication. In our approach, the electronic structure of matter is solved using the self-consistent-charge density functional tight binding (DFTB-SCC) approach, with each CPU responsible for propagating a distinct portion of the macroscopic system. We find that dipole derivatives (Born charges), required for propagating the dynamics, fluctuate significantly during propagation and cannot be
replaced with their time-independent values (i.e., partial charges), as is commonly assumed when using classical force-fields, for achieving quantitative accuracy when propagating the cavity-coupled dynamics. Interestingly, we find that for computing qualitatively accurate linear spectra, the computationally cheaper Mulliken charges can be used in place of Born charges, especially when the nonlinearity in the light-matter interaction term is not substantial. However, such a replacement should not be made for analyzing cavity-modified energy transport properties or chemical reactivity as this replacement leads to a significant heating of the system, as expected.

Using this framework, we demonstrate proof-of-principle simulations of a system containing over 8000 atoms, coupled to an ensemble of cavity modes. We present angle-resolved infrared spectra of liquid water under vibrational strong coupling, targeting both the asymmetric stretch and bending modes. We anticipate that this computational tool will allow for {\it in silico} experiments providing new opportunities to harness confined vacuum radiations in optical cavities to modulate and catalyze chemical reactivity. Based on our computational framework we develop a computational package, namely ``\emph{CavOTF}"\cite{cavOTF} , which is publicly available on our GitHub repository.

{\footnotesize 

\section{Theory}

To demonstrate the feasibility of our on-the-fly approach for simulating vibrational polaritons beyond the long-wavelength approximation, we adapt a modified Holstein-Tavis-Cummings Hamiltonian. Following prior works~\cite{jasrasaria2025jcp, tichauer2021multi, li2024mesoscale}, we consider a simple two-dimensional world extending in $x$ and $y$, with cavity quantization in the $y$ direction and the molecular system extending in the $x$ direction. The cavity radiation modes follow the dispersion relation $\omega(k) = c \sqrt{(\pi/L)^2+k^2}$ where $k = \frac{2\pi j}{L_s}$ ($j = 0, \pm 1, \pm 2, ...$) is the in-place wavevector with $L_s$ as the length of the supercell extending in the in-place direction $x$, $L$ denotes the mirror separation and $c$ is the speed of light in the cavity medium. This simplified setup is used to demonstrate the core concept, while the model can be readily extended to three-dimensional molecular systems and to include multiple polarization directions.
To describe the formation and dynamics of vibrational polaritons, we use the following microscopic light–matter Hamiltonian, formulated in the dipole gauge beyond the long-wavelength approximation~\cite{mandal2023NanoLetters, jasrasaria2025jcp, benson2012comparison} and expressed in mass-weighted coordinates within atomic units ($\hbar = 1 $) as:
{\footnotesize
\begin{align}
    \hat{H}_\mathrm{LM} = \hat{H}_\mathrm{M} + \sum_{k} \omega_k &\left( \hat a_k^\dagger- \sum_{n} {f}_{k}^*(R_n)  \hat M_n\right)\nonumber \\
    \times&\left( \hat a_k - \sum_{n} f_k(R_n)  \hat M_n \right) .
    \label{eq:hlm_bigin}
\end{align}
}


Here, the operators $\hat a_k^\dagger$ ($\hat a_k$) denote the photon creation (annihilation) operator corresponding to the $k$-th cavity mode with frequency $\omega_k$. Further, \(\hat H_\mathrm{M}\) denotes the bare matter Hamiltonian, written as a sum over \(N\) independent cells located at \({R}_n = n \cdot a\) ($a$ is the lattice constant), each containing the same number of molecules. Following recent work~\cite{li2024mesoscale}, we treat the cells as non-interacting, and denote the total dipole operator of the cell \(n\) by \(\hat{M}_n\). The total matter Hamiltonian can be written as

\begin{align}
    \hat{H}_\mathrm{M} = \sum_{n} \hat{H}_{\mathrm{M}_n} = \sum_{n, j} \hat{T}_{n,j} + \hat{V}_{n}(\{r_{n,j}\})
    \label{eq:hm}
\end{align}

where $\hat{T}_{n,j}$ is the nuclear kinetic energy operator for the $j$th atom in the $n$th box and $\hat{V}(\{r_{n,j}\})$ represents potential energy surface of the molecular subsystem. The light-matter coupling enters through the projected dipole operator, $\hat{M}_n = ({\hat{\epsilon} \cdot  \hat \mu_n} )$,  where $\hat \mu_n$ is the total dipole operator of $n$th box and  $\hat{\epsilon} = \hat{x}$ is the  polarization direction of the cavity field which has been approximated to be independent of the in-plane wave-vector $k$ since the polariton spectra is probed near $k \rightarrow 0$.


For a Fabry-P\'{e}rot cavity, the spatial dependence of the cavity field at the cell positions $R_n $ is characterized by the mode function
\begin{equation}
    f_k(R_n) = \frac{g_0}{\sqrt{\omega_k}} e^{i k R_n},
\end{equation}
where $g_0$ is the light–matter coupling constant. In atomic units, it can be written as  $g_0 = \sqrt{\frac{1}{\epsilon_0 \epsilon_r V } } \hat{\epsilon } $ where $V$ is the quantization volume.  The light-matter Hamiltonian in Eq.\ref{eq:hlm_bigin} can be more explicitly written as

\begin{align}\label{eq:hlm2}
  \hat H_\mathrm{LM} &= \hat H_\mathrm{M} + \sum_{k} \omega_k \hat a_k^\dagger \hat a_k
    +  g_0^2 \sum_{m,n,k}  e^{ik(R_n-R_m)}  \hat{M}_m\hat{M}_n \nonumber\\
  &\quad + g_0 \sum_{n,k} \sqrt{\omega_k}\,\big( \hat a_k^\dagger e^{-ikR_n} + e^{ikR_n} \hat a_k\big)\hat{M}_n.
\end{align}
 

This Hamiltonian is schematically illustrated in Fig.~\ref{Fig1}a. According to Eq.~\ref{eq:hlm2}, all photonic modes interact with all molecular boxes. Although the full light–matter system described with Eq.~\ref{eq:hlm2} can be propagated with a parallelized algorithm on a typical high-performance computing cluster (HPC), we pursue a  real-reciprocal-space propagation scheme that allows for a straightforward (lightly-communicating) hub-and-spoke parallelization using Python’s \texttt{socketserver}.  


Since the light-matter coupling term plays a role only around $k \rightarrow 0$, we make the simplifying approximation $\sqrt{\omega_k} (g_0 .\mu_n) \approx \sqrt{\omega_0} (g_0 .\mu_n)$. The photon creation and annihilation operators are then transformed into real-space field operators using $\hat{a}_n =\frac{1}{\sqrt{N}} \sum_{k} e^{ik \cdot R_n}\hat{a}_{k}$.~\cite{jasrasaria2025jcp} Note that  $\sum_k e^{ik\cdot(R_j - R_{j'})} = N\,\delta_{jj'}$. This transformation localizes the interaction in real space (illustrated in Fig.~\ref{Fig1}b) and the total Hamiltonian is then expressed as

\begin{equation}\label{eq:hlm3}
       \hat H_\mathrm{LM} = \hat H_\mathrm{M} + \sum_{k} \omega_k \hat a_k^\dagger \hat a_k  +   \eta \sum_{n} \hat q_n \hat{\mu}_n  +  \frac{\eta^2}{2 \omega_0^2} \sum_{n}  \hat{\mu}_n^2   
\end{equation} 
where $\eta = \sqrt{2N} g_0\omega_0 $ denotes the collective light–matter coupling strength, and $\hat{q}_n = \frac{1}{\sqrt{2\omega_0}}(\hat{a}^{\dagger}_{n} + \hat{a}_{n})$ is the real space photonic coordinate.

In this work, we propagate the dynamics of the nuclear and photonic degrees of freedom classically, i.e. $\{\hat r_{n,j}, \hat q_n \} \rightarrow \{ r_{n,j}, q_n \}$. The equations of motion for both the nuclear and photonic coordinates are derived from the Hamilton’s equations of motion and integrated in time using a velocity-verlet-like scheme. The forces $F_{n,j}^{\mathrm{m}}$ and $F_{n,j}^{\mathrm{c}}$ acting on the nuclear and photonic degrees of freedom, respectively, are written as
\begin{align}\label{eq:dpj}
    F_{n,j}^{\mathrm{m}}  &= -  \nabla_{n,j}  {V}_{n}(\{r_{n,j}\}) - \eta q_n \frac{d \mu_n}{dr_{n,j}} - \frac{\eta^2\mu_n}{\omega_0^2} \frac{d\mu_n}{dr_{n,j}},\\
       F_{n}^{\mathrm{c}}  &= -\eta \mu_n - \frac{d}{d q_{n}}\Bigg(\frac{1}{2}\sum_{k} \tilde{p}^2_k +\omega_k^2 \tilde{q}^2_k\Bigg),
\end{align}
where $\tilde{q}_k = \frac{1}{\sqrt{2\omega_k}}(\hat a_k^\dagger +\hat a_k)$ and $\tilde{p}_k = i\sqrt{\frac{\omega_k}{2}}(\hat a_k^\dagger - \hat a_k)$. Here, we compute the term $-  \nabla_{n,j}  \hat{V}_{n}(\{r_{n,j}\})$ on-the-fly using the DFTB-SCC approach. We integrate the equation of motion of the photonic and nuclear degrees of freedoms using the following scheme that uses a server-client (hub-and-spoke) architecture (schematically illustrated in Fig.~\ref{Fig1}c):

{\bf Step 0.} {\it Initialization.} The matter degrees of freedoms $\{r_{n,j}(0), p_{n,j}(0)\}$ at $t = 0$ are obtained from a long (150 ps) NVT DFTB trajectory.  The photonic degrees of freedom $\{\tilde{q}_{k}, \tilde{p}_{k}\}$ are then initialized performing a constrained molecular dynamics simulation of the Hamiltonian, $H_{LM}(\{r_{n,j} = r_{n,j}(0),~p_{n,j} = p_{n,j}(0)\})$ using a simple velocity-verlet algorithm where $\{r_{n,j}, p_{n,j}\}$ are frozen. 

{\bf Step 1.} {\it Real-space propagation on clients.} We evaluate $-  \nabla_{n,j}  \hat{V}_{n}(\{r_{n,j}\})$, $\mu_n$, $d\mu_n/dr_{n,j}$ (computed numerically, and updated every three steps). Using these quantities, the nuclear degrees of freedom are fully evolved over a single time step $\delta t$, while the photonic momenta are only partially evolved, with their evolution occurring solely due to the localized light–matter coupling term. Specifically, we update the nuclear and photonic degrees of freedom as,
 
\begin{align} \label{eq:pj1}
    \tilde{p}_{n}'\Big(t+\frac{\delta t}{2}\Big) &= \tilde{p}_{n}(t) +  \frac{\delta t}{2}F_{n}^{\mathrm{c}}(\{r_{n,j}(t), q_{n}(t)\})  \\
    p_{n,j}\Big(t+\frac{\delta t}{2}\Big) &= p_{n,j}(t) +  \frac{\delta t}{2}F_{n,j}^{\mathrm{m}}(\{r_{n,j}(t), q_{n}(t)\})   \\
    r_{n,j}(t+\delta t) &= r_{n,j}(t) + p_{n,j}\Big(t+\frac{\delta t}{2}\Big) \delta t \\
      p_{n,j} (t+ {\delta t}  ) &= {p}_{n,j}\Big(t+\frac{\delta t}{2}\Big)  \nonumber\\
      &~~+  \frac{\delta t}{2}F_{n,j}^{\mathrm{m}}(\{r_{n,j}(t + \delta t), q_{n}(t)\}) \\
    \tilde{p}_{n}'(t+{\delta t}) &= \tilde{p}_{n}'\Big(t+\frac{\delta t}{2}\Big) \nonumber\\
    &~~+  \frac{\delta t}{2}F_{n}^{\mathrm{c}}(\{r_{n,j}(t+\delta t), q_{n}(t)\}) 
\end{align}

{\bf Step 2.} {\it Receive photonic coordinates from client and perform fourier transformation on server.} In this step, the server receives all photonic variables $\{ q_n,\tilde{p}_n' \}$ from all the client CPUs and evaluates the Fourier transformed reciprocal-space photonic variables $\{ \tilde{q}_k',\,\tilde{p}_k' \}$ using the following expression 

\begin{align}\label{eq:qk_nk}
\tilde{q}_k'(t+\delta t) = \frac{1}{\sqrt{2\omega_k}} (a_k + a_k^*)\\
\tilde{p}_k'(t+\delta t) = i \sqrt{\frac{\omega_k}{2}} (a_k^* - a_k)
\end{align}
where we introduce the classical complex variables $\{a_k\}$ (and their complex conjugates $\{a_k^*\}$) which are obtained as 

\begin{align}\label{eq:qkink_ak}
& a_k = \sum_{n=0}^{N-1} \sqrt{\frac{\omega_0}{2 N}} \Big( q_n(t) +\frac{i\tilde{p}_{n}'(t+\delta t)}{\omega_0}  \Big) e^ {-ik \cdot R_n }.
\end{align}

{\bf Step 3.} {\it  Photonic propagation in the reciprocal space on  server.}
The photonic variables $\{ \tilde{q}_k,\,\tilde{p}_k \}$ are propagated analytically using
\begin{align}\label{eq:qknk}
    \tilde{q}_k(t+\delta t) = & \tilde{q}_k'(t+\delta t)~ cos(\omega_k \delta t) \nonumber \\
    &+ \tilde{p}_k'(t+\delta t)~sin(\omega_k \delta t)/ \omega_k\nonumber \\
    \tilde{p}_k(t+\delta t) = & \tilde{p}_k'(t+\delta t)~ cos(\omega_k \Delta t) -\omega_k  \nonumber \\
      &~ \tilde{q}_k'(t+\delta t)~sin(\omega_k \delta t).
\end{align}

{\bf Step 4.} {\it Perform inverse Fourier transformation on server and broadcast photonic coordinates to clients.}
Time propagated $k$-space photonic variables $\{ \tilde{q}_k,\,\tilde{p}_k \}$ are inverse Fourier transformed to obtain real space photonic variables $\{ q_n,\tilde{p}_n \}$ using the following expressions

\begin{figure}[h!]
    \centering
    \includegraphics[width=1.0\linewidth]{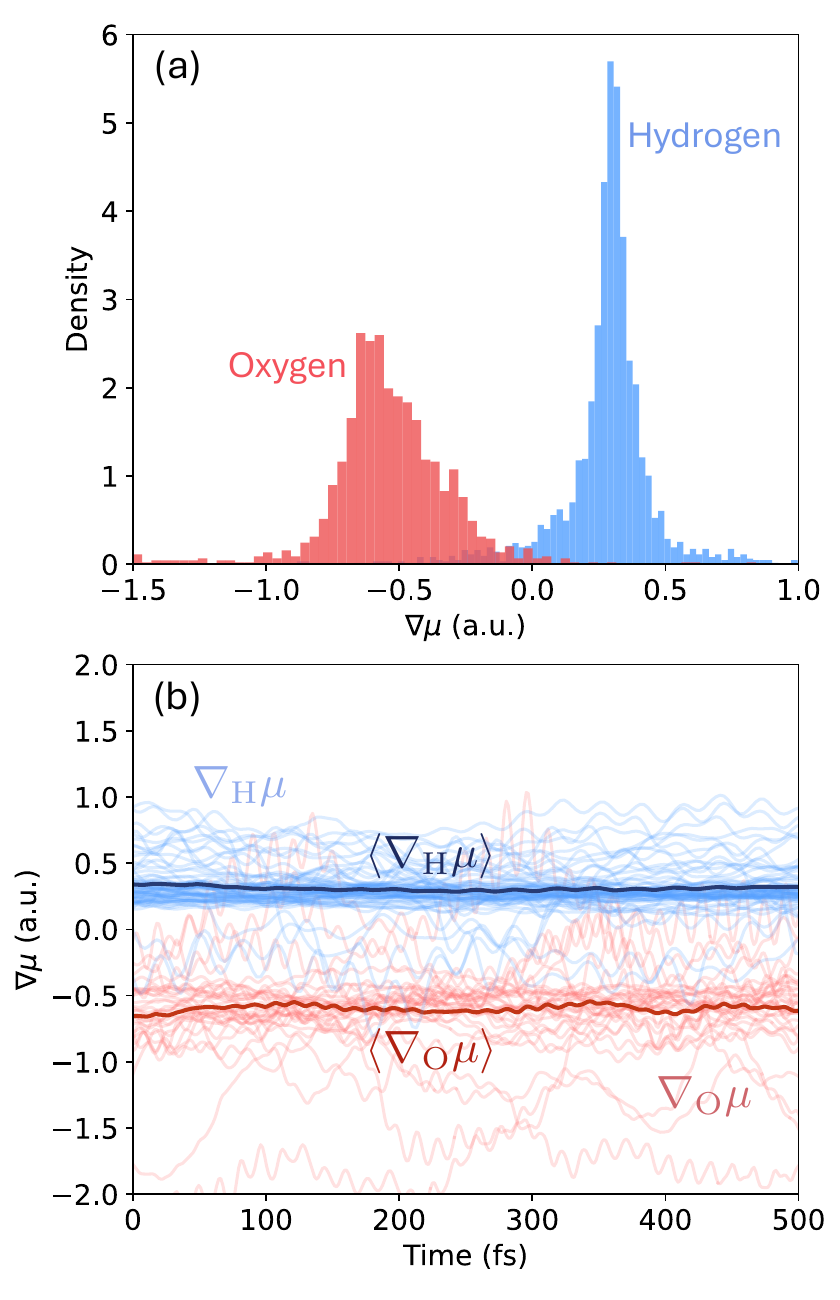} 
    \caption{\footnotesize (a) Born charges distribution of Oxygen and Hydrogen atoms in liquid water and (b) their time-dependent fluctuations. } 
    \label{Fig2}
\end{figure}

\begin{align}\label{eq:nk_qk}
q_n(t+\delta t) &= \frac{1}{\sqrt{2\omega_0}} (a_n + a_n^*)\\
\tilde{p}_n(t+\delta t) &= i \sqrt{\frac{\omega_0}{2}} (a_n^* - a_n)
\end{align}
where $\{a_n\}$ (and their complex conjugates $\{a_n^*\}$) are obtained as 

\begin{align}\label{eq:qkink_an}
& a_n = \sum_{k=0}^{N-1} \sqrt{\frac{\omega_k}{2 N}} \Big( \tilde{q}_k(t+\delta t) +\frac{i\tilde{p}_{k}(t+\delta t)}{\omega_k}  \Big) e^ {ik \cdot R_n }.
\end{align}

The updated values are then sent to all clients. Each client receives only its corresponding pair $(q_n,\tilde{p}_n)$.

{\bf Step 5.} {\it Repeat.}
Continue from {\bf Step 1} for the next time step.


{\bf  Computational details.} Molecular dynamics simulations were performed using the density functional tight binding (DFTB) method with second-order self-consistent charge (SCC) correction. This semiempirical quantum mechanical approach provides a balance between accuracy and computational efficiency, enabling simulations of systems containing hundreds of atoms over picosecond timescales. \cite{elstner2014density,gaus2011dftb3,zheng2009implementation,wickramasinghe2023jpcb,jakowski2012optimization}

In this work, for each $\hat{H}_{M_n}$, a periodic cubic simulation cell of size $10~{\text{\AA}}$  was constructed, containing 33 water molecules to match the experimental density of bulk liquid water. Periodic boundary conditions were applied in all three spatial directions. Canonical (NVT) molecular dynamics simulations were carried out at 300 K using a Nosé–Hoover thermostat and a time step of 0.5 fs.\cite{martyna1996explicit} Each trajectory was propagated for 150 ps, and atomic coordinates were recorded every 1 fs. Five independent NVT trajectories were generated for statistical sampling. From the equilibrated portion of each trajectory (after 80 ps), 100 configurations with velocities were randomly selected. These were then used to initiate microcanonical (NVE) simulations, each propagated for $\sim$2 ps with the same time step of 0.3 fs. Electronic structure calculations during the simulations were performed using SCC-DFTB with Brillouin zone integration approximated via $\Gamma$-point sampling of a 3×3×3 folded supercell, implemented using the `SupercellFolding' scheme in DFTB+.\cite{hourahine2020dftb+}


\section{Results and Discussion}

In this work, the primary computational bottleneck arises from evaluating the derivative of the dipole moment $d\mu_n/dr_{n,j}$ (Born charges), with respect to nuclear motion. In typical classical force fields, atomic partial charges are fixed parameters; consequently, the Born charges reduce to a time-independent constant value $d\mu_n/dr_{n,j} \approx z_{n,j}$. In our DFTB-based simulations, atomic charges are time-dependent and fluctuate along the trajectory.  In Fig.~\ref{Fig2}b we show the time-dependent Born charges on the O and H atoms, labeled as $\nabla_{\mathrm{O}}\mu$ and $\nabla_{\mathrm{H}}\mu$, respectively. We calculate these Born charges using the simple forward finite difference approach, i.e., $\nabla\mu_{n,j} \approx \dfrac{\mu_n(r_{n,j}+\delta r_{n,j}) - \mu_n(r_{n,j})}{\delta r_{n,j}}$. 

Fig.~\ref{Fig2}a presents the Born charge distribution for the O and H atoms sampled from a dynamical evolution of 500 fs. We observe a broad distribution of O's Born charges, with the majority lying between -1 and 0 with a mode near -0.6. H atoms show a much sharper distribution with its peak appearing near +0.3 and most values distributed between -0.2 and +0.8. 

\begin{figure*}[h!]
    \centering
    \includegraphics[width=1.0\linewidth]{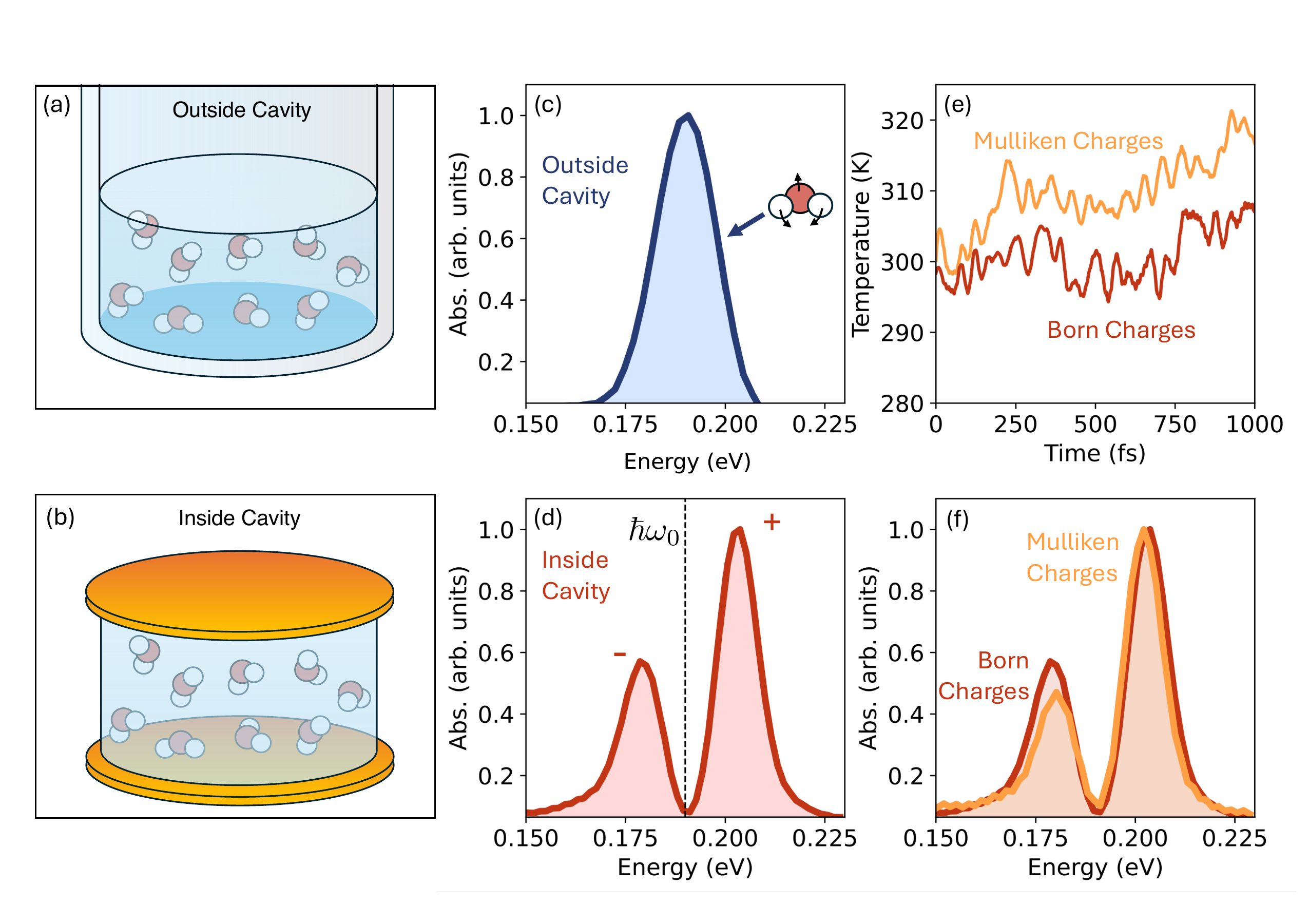} 
    \caption{\footnotesize Schematic figure of  water (a) outside and (b) inside an optical cavity. Simulated molecular spectra spectra of H-O-H bending motion (c) outside and (d) inside of the cavity under vibrational strong coupling.  The effects of using Mulliken vs Born charges on (e) temperature and (f) cavity modified molecular spectra.} 
    \label{Fig3}
\end{figure*}
In Fig.~\ref{Fig2}b we present time-dependent Born charges as well as their statistical averages $\langle \nabla_\mathrm{O} \mu\rangle$ and $\langle \nabla_\mathrm{H} \mu\rangle$. Clearly, the time-dependent fluctuation are large. These large variations illustrate a possible limitation of the presently employed bi-linear (with linear dipole moment) light-matter couplings in prior theoretical works, which have had limited success in explaining cavity modified ground-state chemical kinetics in the collective coupling regime.~\cite{lindoy2024Nanophotonics, campos2023jcp}  Neglecting the fluctuations in the Born charges, underestimates the strong anharmonic nature of molecular vibrations, misrepresenting the actual light–matter interaction under VSC. While outside the scope of the present work, our computational tool enables investigation of non-linear and anharmonic effects in cavity modification of ground-state chemical reactivity, which will be pursued in our future work.

In this work, we study the modification of the molecular and photonic spectra in the  VSC regime.  The cavity modified molecular spectra were computed based on linear response theory, where the spectral intensity $I_{\mathrm{m}}(\omega)$ is computed from the dipole moment autocorrelation function:
\begin{align}
I_{\mathrm{m}}(\omega) &\propto \omega^2 \int_0^\infty \mathrm{d}t\, \cos(\omega t)\, \langle \mu_x(t)\mu_x(0) \rangle \nonumber\\
&\approx \omega^2 \Bigg\langle \int_0^\infty \mathrm{d}t\, \cos(\omega t)\mathcal{S}[\mu_{x}(t) \mu_{x}(0) ] \Bigg\rangle 
\end{align}
where $\langle ... \rangle$ denotes statistical averaging, and $\mathcal{S}$ is a smoothening function (a weighted moving average) which allows us to obtain reasonably converged  spectra with a few number of trajectories, as was done in Ref.~\citenum{taoEligit}. We used 40 individual trajectories to  compute the molecular  spectra which couple to single cavity mode.

Fig.~\ref{Fig3}c  shows the simulated molecular spectra featuring a peak corresponding to the H-O-H bending mode in bulk water outside of the cavity (schematically illustrated in Fig.~\ref{Fig3}a). This particular peak appears at $\sim$0.19 eV with a  narrow profile and is in good agreement with past experimental and theoretical works.
~\cite{yu2020NatureComm,carpenter2017jpc} 

In Fig.~\ref{Fig3}d, we illustrate the cavity modified molecular spectra when considering a single radiation mode coupled to a single box of water (99 atoms). The light-matter Hamiltonian for this reduced system is written as 
\begin{equation}\label{eq:hlm4}
       \hat H_\mathrm{LM} = \hat H_\mathrm{M} +  \omega_0 \hat a_0^\dagger \hat a_0  +   \eta   \hat q_0 \hat{\mu}_0  +  \frac{\eta^2}{2 \omega_0^2} \sum_{n}  \hat{\mu}_0^2.  
\end{equation}

\begin{figure*}[h!]
    \centering    \includegraphics[width=1.0\linewidth]{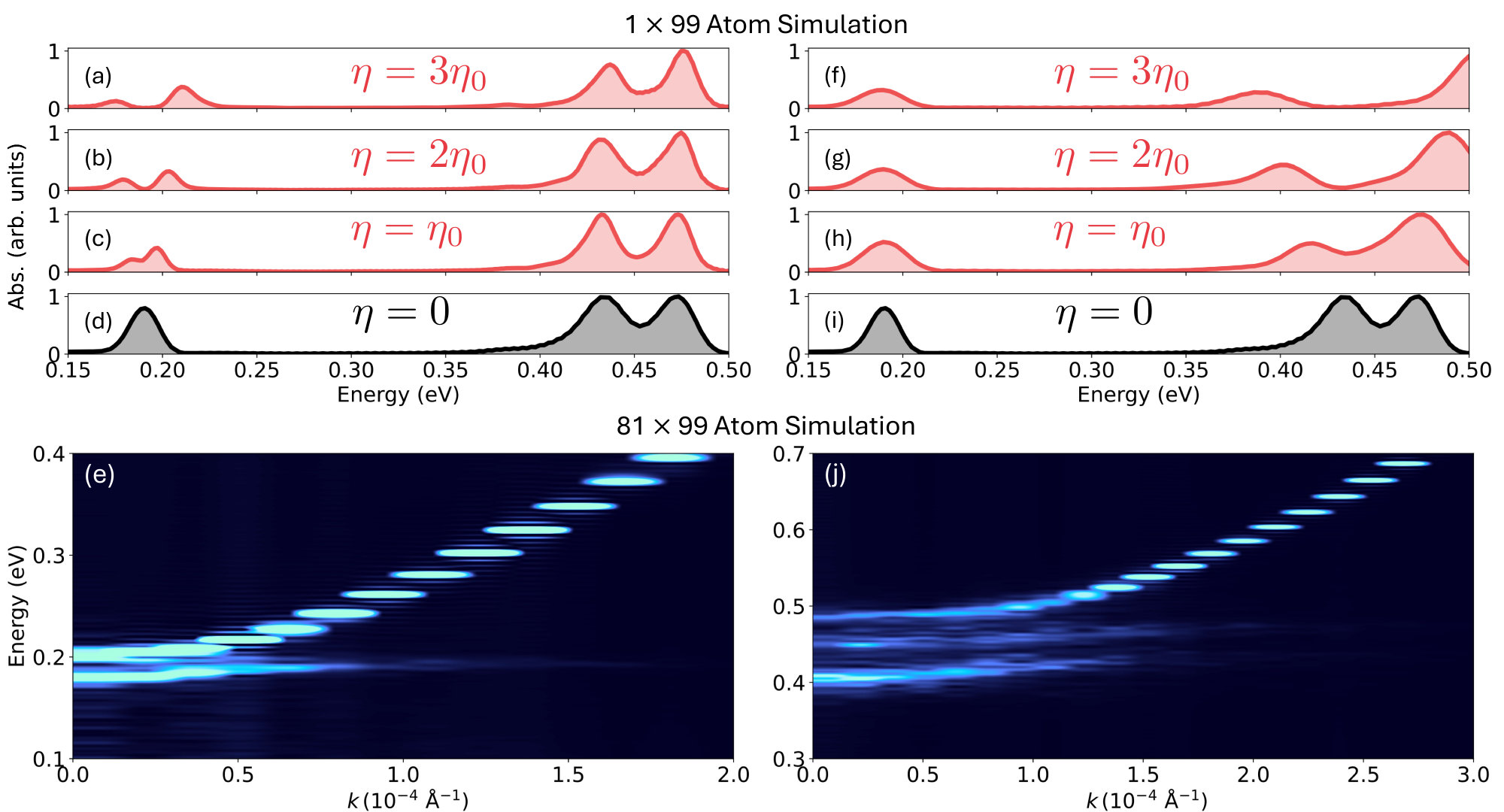}
    \caption{\footnotesize Cavity modified molecular spectra and angle resolved photonic spectra under vibrational strong coupling.  Panels (a)–(d) show cavity modified molecular spectra at various light–matter coupling strengths for $\eta_0 = 0.0001$ with $\omega_0 = 0.19\:\mathrm{eV}$, and panels (f)–(i) show the corresponding spectra for $\eta_0 = 0.00034$ with $\omega_0 = 0.43\:\mathrm{eV}$. Angle resolved photonic spectra are shown in panel (e) for $\omega_0 = 0.19~\text{eV}$ and in panel (j) for $\omega_0 = 0.43~\text{eV}$.} 
    \label{Fig4}
\end{figure*}
Here we have adjusted the mirror spacing to tune the cavity frequency $\hbar \omega_0$ to 0.19 eV, to be in resonance with the H-O-H bending mode. As expected, light-matter interactions lead to a Rabi-splitting  ($\sim$0.024 eV for $\eta = 2 \eta_0 $ a.u. chosen here) with two peaks corresponding to the upper ($+$) and lower ($-$) vibro-polaritons.

In this calculation, we used the computationally expensive Born charges in propagating the dynamics of the light-matter hybrid system. In Fig.~\ref{Fig3}f we explore the possibility of using the computationally inexpensive Mulliken charges as an approximate replacement for the Born charges in order to compute the cavity modified spectra. Interestingly, Fig.~\ref{Fig3}f shows that the Mulliken charges do provide reasonably accurate polaritonic spectra illustrating an efficient route to computing vibro-polariton linear spectra.

However, such an approximation (replacing Born charges with Mulliken) does lead to long-time heating of the system, making it potentially incompatible for studying cavity modified chemical dynamics or energy transport. Fig.~\ref{Fig3}e compares the temperature fluctuations over 1 ps. We observed that the temperature fluctuation of water does not exceed the $\sim$307 K, when the Born charges were used (red curve).  In contrast, simulations using Mulliken charges exhibit a continuous heating: it exceeds the 310 K within 250 fs and continues to rise over 320 K (orange curve).

In Fig.~\ref{Fig4}a-c, we illustrate the effect of total light matter coupling $\eta$ given in Eq.~\ref{eq:hlm3} on molecular spectra. Here, we have tuned the cavity frequency $\hbar\omega$ to be in resonance with the H-O-H bending mode. It shows that as $\eta$ increases from $\eta_0 $ to $3\eta_0$ a.u. (where $\eta_0 = 0.0001$ a.u.), the upper and lower polariton peaks separate, reflecting an increase in the Rabi-splitting from 0.012 eV to 0.036 eV, proportional to the total light matter coupling ($\propto\eta $), as is expected from a simple Jaynes–Cummings model~\cite{mandal2023ChemicalReviews, jasrasaria2025jcp, li2022Annualreview}. This indicates the lack of anharmonic or nonlinear effects in this particular molecular vibration which also explains the success of using the Mulliken charges in this specific instance. 

\begin{figure*}[h!]
    \centering
    \includegraphics[width=1.0\linewidth]{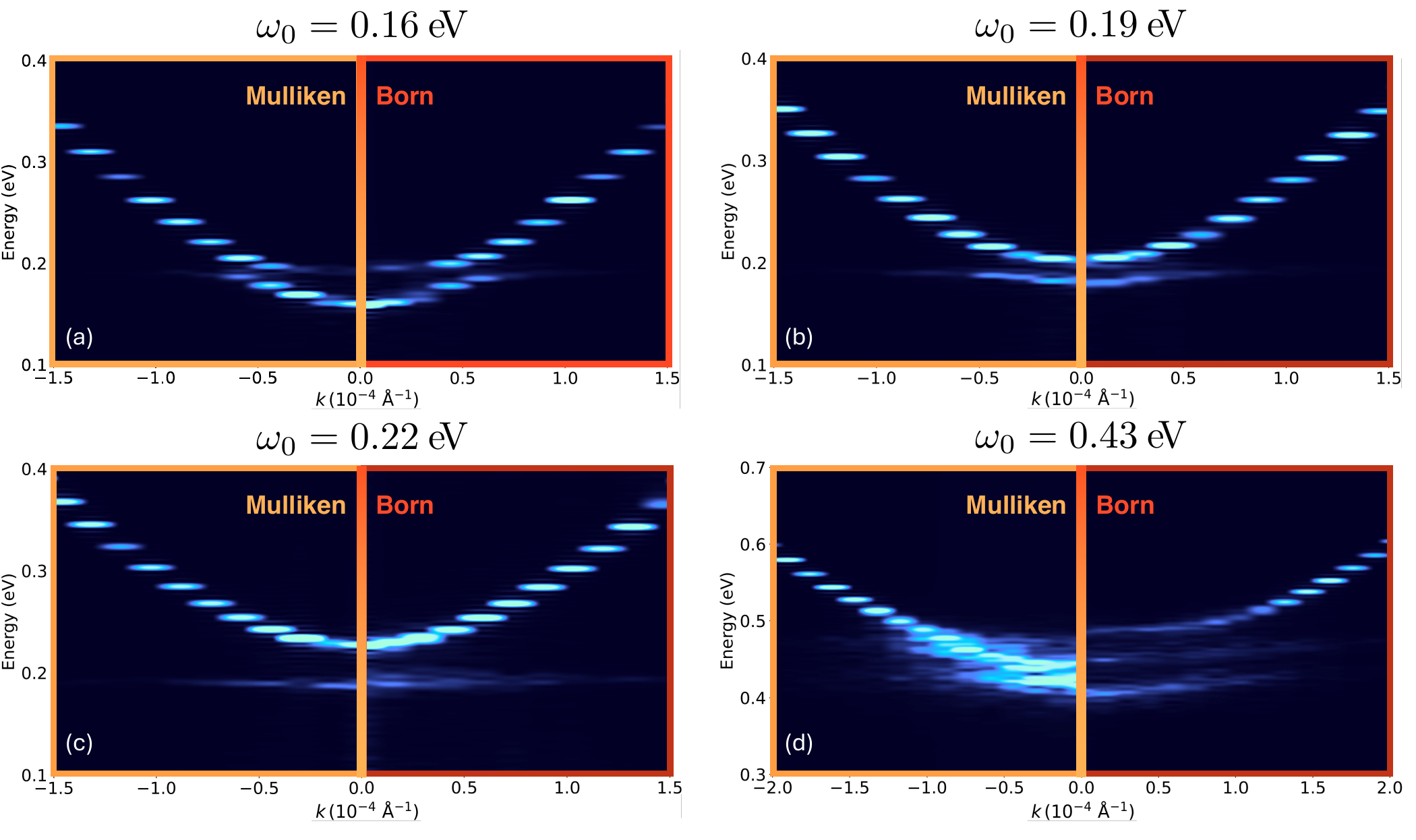} 
    \caption{\footnotesize  Angle-resolved photonic spectra under vibrational strong coupling when using Mulliken charges vs Born charges at (a) $\omega_0 = 0.16$ eV,  $\omega_0 = 0.19$ eV, $\omega_0 = 0.22$ eV, and $\omega_0 = 0.43$ eV. Here we use the light-matter coupling $\eta = \eta_c \cdot \omega_0^{3/2}$ with $\eta_c = 0.00105$.} 
    \label{Fig5}
\end{figure*}



In Fig.~\ref{Fig4}f-h, we show the effect of $\eta$ (chosen as multiples of $\eta_0 = 0.00034$ a.u., a constant) on molecular spectra when tuning the cavity frequency  to 0.43 eV. This frequency is nearly in resonance to the H–O–H symmetric stretching mode that appears at $\sim$ 0.43 eV, in the bare molecular spectrum shown in Fig.~\ref{Fig4}i. In the bare molecular spectrum, the symmetric and asymmetric H–O–H stretching vibrations can be observed at $\sim$ 0.43 eV and $\sim$ 0.47 eV, respectively. Unlike in H-O-H bending mode the vibro-polariton modes formed here combines two molecular vibrations with one cavity radiation mode leading to more complex polariton modified molecular spectra. With an increase in 
$\eta$, from $\eta_0$ to $3\eta_0$ (here $\eta_0 = 0.00034$ a.u.), we observe these  two  peaks moving farther apart.

In addition to these single cavity mode calculations, we now present the angle resolved vibro-polariton spectra which are computed from the photonic autocorrelation function $I_{\mathrm{k}}(\omega)$ which is expressed as
\begin{align}
I_{\mathrm{k}}(\omega) &\propto \omega^2 \int_0^\infty \mathrm{d}t\, \cos(\omega t)\, \langle \tilde{q}_k(t)\tilde{q}_k(0) \rangle \nonumber\\
&\approx \omega^2 \Bigg\langle \int_0^\infty \mathrm{d}t\, \cos(\omega t)\mathcal{S}[\tilde{q}_k(t) \tilde{q}_k(0) ] \Bigg\rangle
\end{align}
where $\tilde{q}_k$ is the photonic position in the reciprocal space. We computed this dispersion using 5 independent trajectories (for which we found to converge the polariton spectra for this specific setup). Each trajectory includes 81 localized cavity radiation modes,  with each mode coupling to a box of water each containing 33 water molecules as schematically shown in Fig.~\ref{Fig1}b.

Fig.~\ref{Fig4}e shows the angle resolved vibro-polariton spectra as a function of the in-plane cavity wave-vector $k$ with the $k = 0$ cavity mode tuned to 0.19 eV. Thus, $k=0$ cavity mode is in resonance with the H–O–H bending vibration, forming the upper and lower polariton bands. When $k$ increases, the cavity photon energy rises while the molecular vibrational energy remains flat. This causes the upper polariton branch to become increasingly photonic, while the lower branch becomes increasingly matter-like at higher $k$.

In Fig.~\ref{Fig4}j, we tuned the cavity mirror spacing to adjust the cavity frequency of the $k = 0$ cavity mode to be 0.43 eV. Here we can clearly see two flat bands, corresponding to symmetric and asymmetric stretching modes simultaneously hybridized with cavity band leading to the formation of three polaritonic bands.




Finally in Fig.~\ref{Fig5}, we compute the angle resolved spectra and benchmark the usage of Mulliken charges as an approximate replacement of Born charges. Here we tune the photon frequency to resonate with  the H-O-H bending (0.19 eV) or H-O-H stretching frequencies (0.43 eV) as well as two off-resonance frequencies $0.16 \: \mathrm{eV}$ and $0.22 \: \mathrm{eV}$.  Fig.~\ref{Fig5}a-c show that the spectra computed using Mulliken charges closely match those obtained with Born charges, capturing all qualitative features of the spectra. In contrast, in Fig.~\ref{Fig5}(d) we observe noticeable deviations in the angle resolved spectra when Born charges are replaced with Mulliken charges. This is because the cavity modes are simultaneously coupling to the H-O-H symmetric and asymmetric stretching which are known to exhibit substantially more nonlinearities in comparison to the bending motion~\cite{vinaykin2012vibrational}. This illustrates that in the presence of substantial molecular nonlinearity the use of Mulliken charges is insufficient and should be approached with caution.

\section{Conclusion} 
In this work, we propagate the dynamics of strongly coupled vibro-polaritons, formed by coupling molecular vibrations with confined radiation modes inside an optical cavity, where molecular potentials, dipoles and their gredients are computed on-the-fly. In our approach, we adapt a real-space description of the light–matter interaction and exploit its sparsity to devise a low-communication, hub-and-spoke server–client parallelization scheme. Using this approach, we develop an open-source code CavOTF\cite{cavOTF} which is available on GitHub. 

We illustrate the utility of our approach by computing the cavity modified molecular and angle resolved photonic spectra. Obtaining the dipole gradients, namely the Born charges, is the computational bottleneck of our approach as they are evaluated numerically. We find that Mulliken charges can often be used as a computationally inexpensive replacement for computing qualitatively accurate linear spectra. However, when substantial molecular nonlinearity persists, this approach produces less accurate linear spectra. As expected, using Mulliken charges leads to spurious heating of the system and therefore should not be used when studying cavity modified chemical dynamics or cavity modified energy transport.

\section{Acknowledgments}
This work was supported by the Texas A\&M startup funds and the Strategic Transformative Research Program (STRP) grant provided by the Texas A\&M University. This work used TAMU FASTER and ACES at Texas A\&M University through allocation PHY230021, TAMU LAUNCH at Texas A\&M University, and SDSC Expanse at the San Diego Supercomputer Center through allocations CHE250156 and CHEM250162 from the Advanced Cyberinfrastructure Coordination Ecosystem: Services\cite{boerner2023ACCESS} \& Support (ACCESS) program, which is supported by U.S. National Science Foundation grants \#2138259, \#2138286, \#2138307, \#2137603, and \#2138296. A.M. appreciates discussion with Xin Yan, Gerrit Groenhof, Pengfei Huo and Elious Mondal. The authors thank Michael Fowler for his assistance in preparing Fig. 3.

\bibliography{bib.bib}

 }
\end{document}